\documentclass{ws-procs9x6}
\begin{document}

\title{Y-type Flux-Tube Formation in Baryons}

\author{Toru~T.~Takahashi}

\address{Yukawa Institute for Theoretical Physics,\\
Kyoto University,
Sakyo, Kyoto 606-8502, Japan\\ 
E-mail: ttoru@yukawa.kyoto-u.ac.jp}

\author{Hideo~Suganuma and Hiroko Ichie}

\address{Faculty of Science, Tokyo Institute of Technology,\\
Ohokayama 2-12-1, Tokyo 152-8551, Japan}


\maketitle

\abstracts{
For more than 300 different patterns of the 3Q systems, 
the ground-state 3Q potential $V_{\rm 3Q}^{\rm g.s.}$
is investigated using SU(3) lattice QCD with 
$12^3\times 24$ at $\beta=5.7$ and
$16^3\times 32$ at $\beta=5.8, 6.0$ at the quenched level.
As a result of the detailed analyses, 
we find that the ground-state potential $V_{\rm 3Q}^{\rm g.s.}$
is well described with so-called Y-ansatz as
$
V_{\rm 3Q}=-A_{\rm 3Q}\sum_{i<j}\frac1{|{\bf r}_i-{\bf r}_j|}
+\sigma_{\rm 3Q} L_{\rm min}+C_{\rm 3Q},
$
with the accuracy better than 1\%.
Here, $L_{\rm min}$ denotes the minimal value of total flux-tube length. 
We also study
the excited-state potential $V_{\rm 3Q}^{\rm e.s.}$
using lattice QCD with $16^3\times 32$ at $\beta=5.8, 6.0$ 
for more than 100 patterns of the 3Q systems.
The energy gap between $V_{\rm 3Q}^{\rm g.s.}$ and $V_{\rm 3Q}^{\rm e.s.}$, 
which physically means the gluonic excitation energy, 
is found to be about 1 GeV 
in the typical hadronic scale.
Finally, we suggest a possible scenario which connects the
success of the quark model to QCD.
\vspace{-0.3cm}
}

\section{Introduction}

It is widely accepted that Quantum Chromodynamics (QCD) is the
underlying theory of the strong interaction for hadrons and nuclei.
In a spirit of the particle physics, 
it is strongly desired to describe the hadron dynamics
directly from QCD.
However, due to the strong coupling nature of QCD in the infrared region,
it still remains a big challenge to derive physical quantities
directly from QCD in an analytic manner.

For the nonperturbative analysis of QCD, lattice QCD Monte Carlo
calculations have been established as a reliable method
directly based on QCD.
Lattice QCD calculations enable us to perform a 
nonperturbative quantitative analysis in a
model-independent way.
For instance, lattice QCD calculations have been very successful
in reproducing hadron mass spectra.\cite{CPPACS01}
%
%
Nevertheless, these results hardly inform us the internal structure of hadrons.
It would be also indispensable to comprehend hadrons 
directly from quarks and gluons, the degrees of freedom of QCD.

For this aim, we cast light on the inter-quark potentials,
which are directly responsible for the hadron properties
and their internal structures at the quark-gluon level.

\vspace{-0.2cm}

\section{Ground-state three quark potential}

For the quark-antiquark (Q$\rm\bar Q$) potential, which 
is one of the most fundamental quantities in QCD,
many lattice QCD studies have been performed.
As a result, Q$\rm\bar Q$ potential is known to be reproduced
by a simple sum as 
$
V_{\rm Q\bar Q}=
-\frac{A_{\rm Q\bar Q}}{r}+\sigma_{\rm Q\bar Q} r+C_{\rm Q\bar Q}
$
with $r$ the distance between quark and antiquark.
Here, $\sigma\simeq 0.89$ GeV/fm denotes the string tension
in the flux-tube picture.

However, in contrast with a number of lattice QCD studies 
for the Q$\rm\bar Q$ potential, 
almost no lattice studies were done before our study\cite{TMNS,TSI} 
for  the three-quark (3Q) potential $V_{\rm 3Q}$, 
which is directly responsible for the baryon properties.
We study the ground-state 3Q potential on more than 300 patterns
of the spatial quark configurations,
using SU(3) lattice QCD at the quenched level
with $12^3\times 24$ at $\beta=5.7$ and with 
$16^3\times 32$ at $\beta=5.8, 6.0$.\cite{TMNS,TSI}

\begin{figure}[ht]
\vspace{-0.4cm}
\centerline{\epsfxsize=6.1cm{\epsfbox{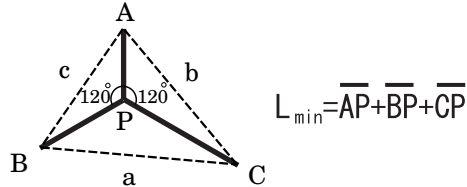}}}
\vspace{-0.2cm}
\caption{The flux-tube configuration in the
3Q system with the minimal length of the Y-type flux-tube, $L_{\rm min}$.
\label{lmin}}
\end{figure}

As a result, we find that the 3Q potential $V_{\rm 3Q}$
can be reproduced by so-called Y-ansatz as~\cite{TMNS,BPV95}
\begin{equation}
V_{\rm 3Q}=-A_{\rm 3Q}\sum_{i<j}\frac1{|{\bf r}_i-{\bf r}_j|}
+\sigma_{\rm 3Q} L_{\rm min}+C_{\rm 3Q},
\label{yansatz}
\end{equation}
with the accuracy better than 1\% for all the different patterns of 
quark configurations.
Here, $L_{\rm min}$ denotes the minimal value of total flux-tube
length,\cite{KS75,CI86} 
which is schematically illustrated in Fig.\ref{lmin}.

In the fit analysis with Y-ansatz, we find two remarkable features\cite{TMNS};
the universality of the string tension as 
$\sigma_{\rm 3Q}\simeq \sigma_{\rm Q\bar Q}$ and 
the one-gluon-exchange consequence as $A_{\rm 3Q}\simeq \frac12 A_{\rm Q\bar Q}$.

Furthermore, we consider also Y-ansatz with Yukawa-type two-body force as
$
V^{\rm Yukawa}_{\rm 3Q}=-A^{\rm Yukawa}_{\rm 3Q}
\sum_{i<j}\frac{\exp (-m_{\rm B}|{\bf r}_i-{\bf r}_j|)}{|{\bf r}_i-{\bf r}_j|}
+\sigma^{\rm Yukawa}_{\rm 3Q} L_{\rm min}+C^{\rm Yukawa}_{\rm 3Q},
$
which is a conjecture from the dual superconductor scenario on the QCD
vacuum. Here, $m_{\rm B}$ denotes the dual gluon mass in the dual
superconductor picture.
However, we find no evidence of the Yukawa-type two-body force\cite{TMNS}
and again confirm the adequacy of the Y-ansatz form in 
Eq.(\ref{yansatz}), {\it i.e.}, 
$\sigma^{\rm Yukawa}_{\rm 3Q}\simeq \sigma_{\rm 3Q}$,
$A^{\rm Yukawa}_{\rm 3Q}\simeq A_{\rm 3Q}$, $m_{\rm B}\simeq 0$.

\begin{figure}[ht]
\centerline{\epsfxsize=11.5cm{\epsfbox{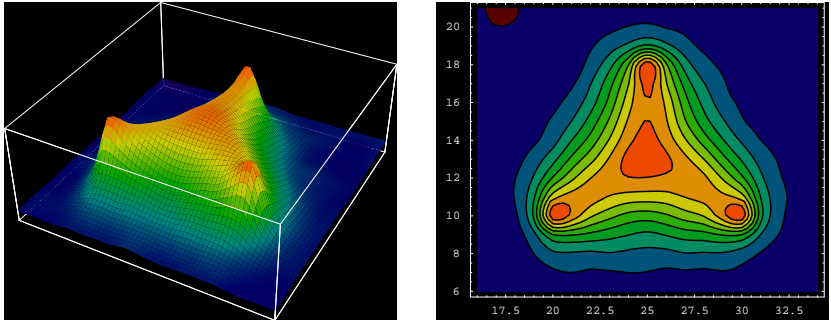}}}
\caption{The flux-tube profile in the spatially-fixed 3Q system,
in the MA projected QCD.$^6$
The distance between the junction and each quark is about 0.5 fm.
\label{fluxprof2}}
\vspace{-0.4cm}
\end{figure}

Finally, as a direct evidence of the Y-type flux-tube formation, 
we show in Fig.\ref{fluxprof2} 
the flux-tube profile in the spatially-fixed 3Q system 
obtained using maximally-abelian (MA) projected lattice QCD.\cite{I02}
The action density in ${\bf R}^3$ is calculated under the presence of spatially-fixed
three quarks in a color-singlet state.
Here, the distance between the junction and each quark is about 0.5 fm.
We thus observe Y-type flux-tube formation in lattice QCD.

\section{Excited-state three quark potential}
\subsection{Lattice QCD study of gluonic excitations}

We also investigate the gluonic excitation in the spatially-fixed 3Q system
using SU(3) lattice QCD.
In the flux-tube picture,\cite{KS75,CI86} gluonic excitation modes would be regarded
as the vibrational modes of a flux-tube.
In the valence picture, they correspond to the
hybrid hadrons such as $q\bar qG$ or $qqqG$.
In particular, hybrid hadrons are probable candidates of 
the exotic hadrons, which has exotic quantum numbers such as
$J^{PC}=0^{--},0^{+-},1^{-+},2^{+-},...$,
which cannot be constructed within a simple quark model.
It is then worth investigating gluonic excitations
directly from QCD, from the experimental viewpoint
as well as the theoretical viewpoint.\cite{CP02}

\begin{figure}[ht]
\vspace{-0.3cm}
\centerline{
\epsfxsize=5.5cm{\epsfbox{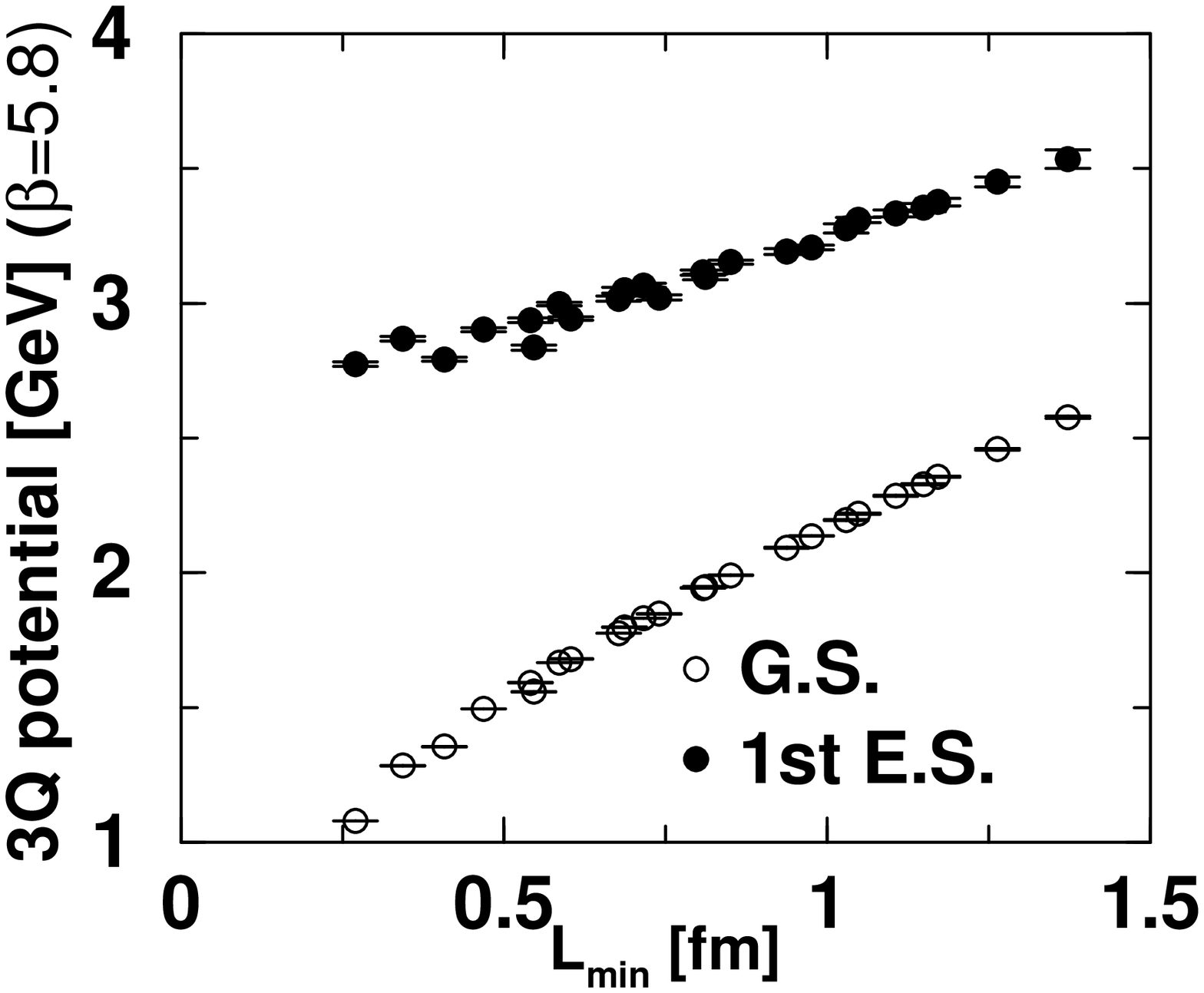}}
\epsfxsize=5.5cm{\epsfbox{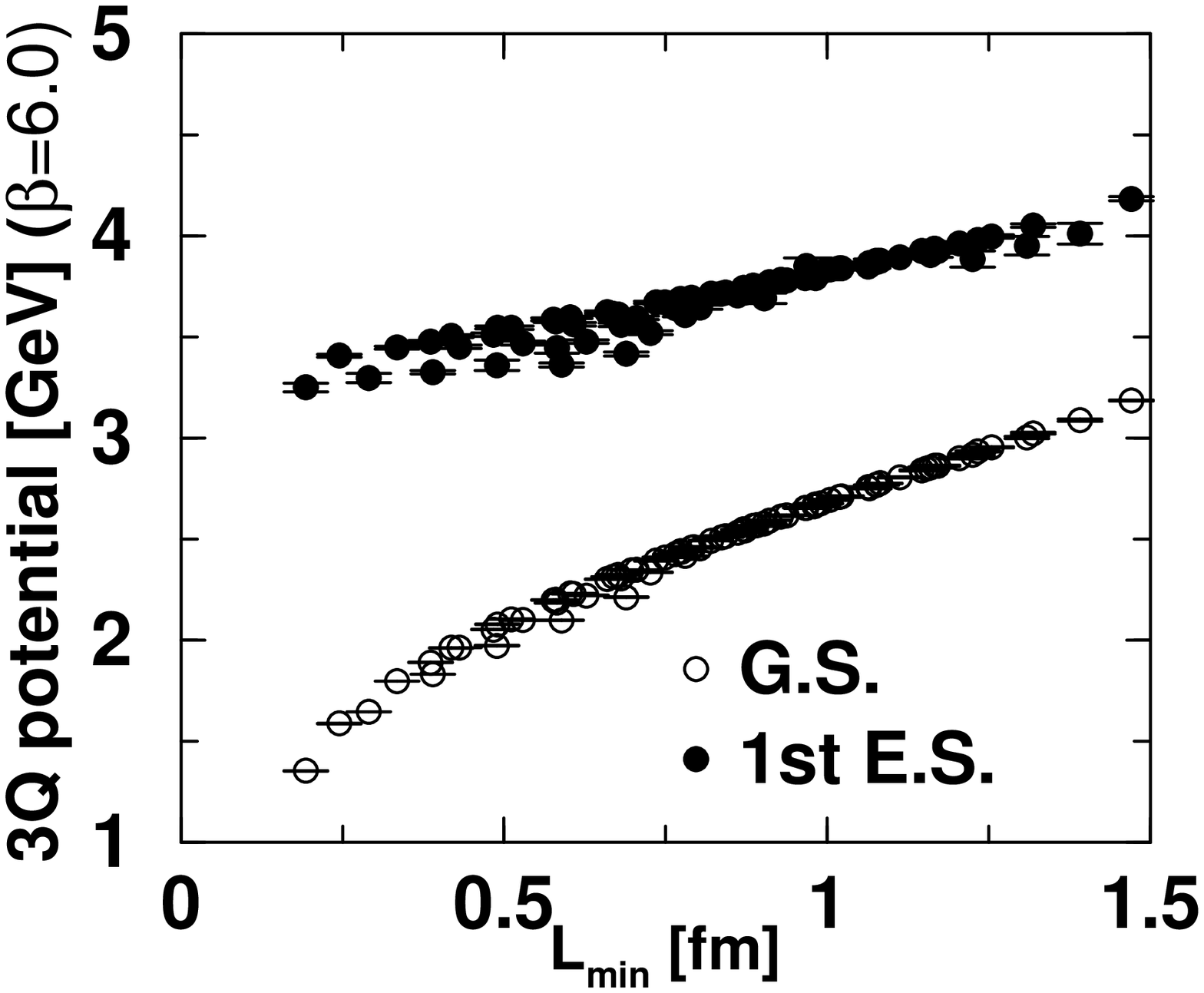}}}
\caption{The lattice QCD results for the ground-state 3Q potential 
$V_{\rm 3Q}^{\rm g.s.}$ (open circles) and the 1st excited-state 3Q
potential $V_{\rm 3Q}^{\rm e.s.}$ (filled circles) as the function
of $L_{\rm min}$.
These lattice results at $\beta=5.8$ and $\beta=6.0$ well coincide 
besides an overall irrelevant constant. 
\label{gluonic}}
\vspace{-0.35cm}
\end{figure}

In Fig.\ref{gluonic} we show the first lattice QCD results for the
excited-state 3Q potential
obtained with $16^3\times 32$ lattice at $\beta =5.8, 6.0$
at the quenched level.\cite{TS03}
Here, the horizontal axis denotes the minimal length of the Y-type flux-tube, 
$L_{\rm min}$, as a label to distinguish the three-quark configuration.
The vertical axis denotes the energy induced
by three static quarks in a color-singlet state.
The open symbols are for the ground-state potential 
$V^{\rm g.s.}_{\rm 3Q}$ 
discussed in the previous section, and the filled symbols are for the excited-state potential
$V^{\rm e.s.}_{\rm 3Q}$in the 3Q system.
The energy difference $\Delta E\equiv V^{\rm e.s.}_{\rm 3Q}-V^{\rm g.s.}_{\rm 3Q}$
physically means the gluonic excitation energy.

We find that the gluonic excitation energy $\Delta E$ 
is more than 1 GeV in the typical hadronic scale as 
$0.5{\rm fm} \leq L_{\rm min}\leq 1.5{\rm fm}$,
which is large in comparison with the excitation energies of quark origin.
(For the Q${\rm \bar Q}$ system, a large gluonic excitation
energy is reported in recent lattice studies.\cite{JKM03})
Such a gluonic excitation would contribute significantly in
the highly-excited baryons with the excitation energy above 1 GeV. in fact, 
the lowest hybrid baryon,\cite{CP02} which is described as $qqqG$ in
the valence picture, is expected to have a large mass of about 2 GeV.
This large gluonic excitation energy also gives us the reason for the
great success of the simple quark model.\cite{TS03}

\subsection{Success of the quark model}

\begin{figure}[ht]
\vspace{-0.3cm}
\centerline{\epsfxsize=9.0cm{\epsfbox{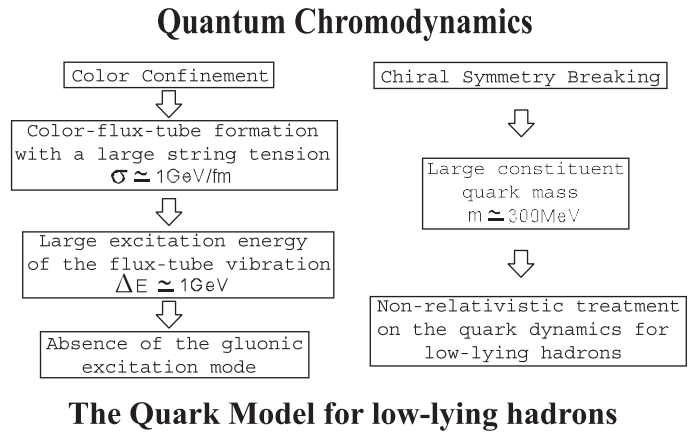}}}
\caption{A possible scenario connecting the success of the simple
quark model to QCD.
\label{behindthesuccess}}
\vspace{-0.7cm}
\end{figure}

The simple quark model contains only constituent quarks as explicit degrees of
freedom, and does not have gluonic modes.
Nevertheless, the quark model has been very successful
in reproducing the low-lying hadron properties and spectra especially for baryons,
in spite of the non-relativistic treatment and 
the absence of gluonic excitation modes.
It is true that the non-relativistic treatment for hadron mass spectra
can be justified by the spontaneous chiral symmetry breaking,
which gives rise to a large constituent quark mass of about 300 MeV,
but we have no reason based on QCD for the absence of the gluonic mode.
Now, directly based on QCD, the gluonic excitation energy $\Delta E$ is found to be 
large, in comparison with quark excitation energy.
We propose a possible scenario connecting the success of the
quark model to QCD\cite{TSI,TS03} in Fig.\ref{behindthesuccess}.

\end{document}